\documentclass[twocolumn,twoside,slac_two]{revtex4}
\usepackage{graphicx}
\usepackage{fancyhdr}
\def\msun{${\rm M}_\odot$ }

\begin{document}

\title{X-ray States of Black-Hole Binaries and Implications for the Mechanism of Steady Jets}


\author{R. A. Remillard}
\affiliation{MIT Kavli Institute for Astrophysics and Space Research \\ Cambridge, MA 02139, USA}

\begin{abstract}
{\it RXTE} and other high-energy observatories continue to probe the
properties of stellar-size black holes and the physics of accretion
using bright X-ray transients in the Galaxy.  Progress has been made
in recognizing that the three states of active accretion are related
to different physical elements that may contribute radiation: the
accretion disk, a jet, and a compact and radio-quiet corona.  Each
X-ray state offers potential applications for general relativity in
the regime of strong gravity. The temporal evolution of X-ray states
is displayed for a few representative black-hole systems.
Radio investigations have shown conclusively that the hard X-ray state is
associated with the presence of a steady radio jet. The three X-ray states 
can be synthesized with the ``unified model for black hole
binary jets'' by Fender, Belloni, \& Gallo (2004) to gain further
insights into the disk:jet connection. The ``jet line''
appears to coincide with the hard limit of the SPL state.
Furthermore there are broad power peaks in PDS that appear
to be confined to intermediate and hard states where a jet is present.
This suggests that broad power peaks exhibit temporal signatures
of non-thermal processes that are related to the jet mechanism,
rather than properties inherent to a standard accretion disk.

\end{abstract}

\maketitle

\thispagestyle{fancy}


\section{X-ray States of Black Hole Binaries}
The X-ray states of black hole binary systems have been re-defined
recently in terms of quantitative criteria that utilize both X-ray
energy spectra and power density spectra (PDS) \cite{mcc05}.  This
effort was undertaken in response to the lessons of extensive
monitoring campaigns with the {\it Rossi} X-ray Timing Explorer ({\it
RXTE}), which revealed the full complexity of spectral and timing
evolution exhibited by accreting black-hole binary systems
\cite{sob99,hom01}. 

The redefinition of X-ray states uses four criteria: $f_{disk}$,
the ratio of the disk flux to the total flux (both unabsorbed) at 2-20
keV; the power-law photon index ($\Gamma$) at energies below any break
or cutoff; the integrated rms power ($r$) in the PDS at 0.1--10 Hz,
expressed as a fraction of the average source count rate; and the
integrated rms amplitude ($a$) of a quasi-periodic oscillation (QPO)
detected in the range of 0.1--30 Hz. PDS criteria ($a$ and $r$)
utilize a broad energy range, and the bandwidth of the {\it RXTE} PCA
instrument for typical X-ray sources is effectively 2--30 keV.

It had been known for decades that the energy spectra of outbursting
black holes often exhibit composite spectra consisting of two
broadband components.  There is a multi-temperature accretion disk
\cite{sha73,mak86,gie04,li05} with a characteristic temperature near 1
keV.  {\bf Thermal state} designations draw attention to those times
when the radiation is dominated by the heat from the inner accretion
disk.  The thermal state (formerly the ``high/soft'' state) is defined
\cite{mcc05} by the following three conditions: (1) the disk
contributes more than 75\% of the total unabsorbed flux at 2--20 keV,
i.e. $f > 0.75$, (2) there are no QPOs present with integrated
amplitude above 0.5\% of the mean count rate, i.e. $a_{max} < 0.005$,
and (3) the integrated power continuum is low, with $r < 0.06$.
 
In principle, the normalization constant for the thermal component may
allow numerical estimates of the radius of the inner accretion disk
($R_{in}$), if the source distance and disk inclination are accurately
known \cite{zha97a,mer00,zim05,li05}. However, such estimates depend
on disk models computed under general relativity (GR), with careful
attention to the inner disk boundary condition and to effects of
radiative transfer \cite{shi95}.  The ongoing efforts to utilize 
GR disk models and consider all forms of systematic problems
in dealing with the inner disk boundary condition may yield
reliable measures of $R_{in}$.  This would lead to estimates
of the black-hole spin parameter for cases where the black-hole mass
is well constrained via measurements of companion star motion in the
binary system.

\begin{figure*}[t]
\centering
\includegraphics[width=135mm]{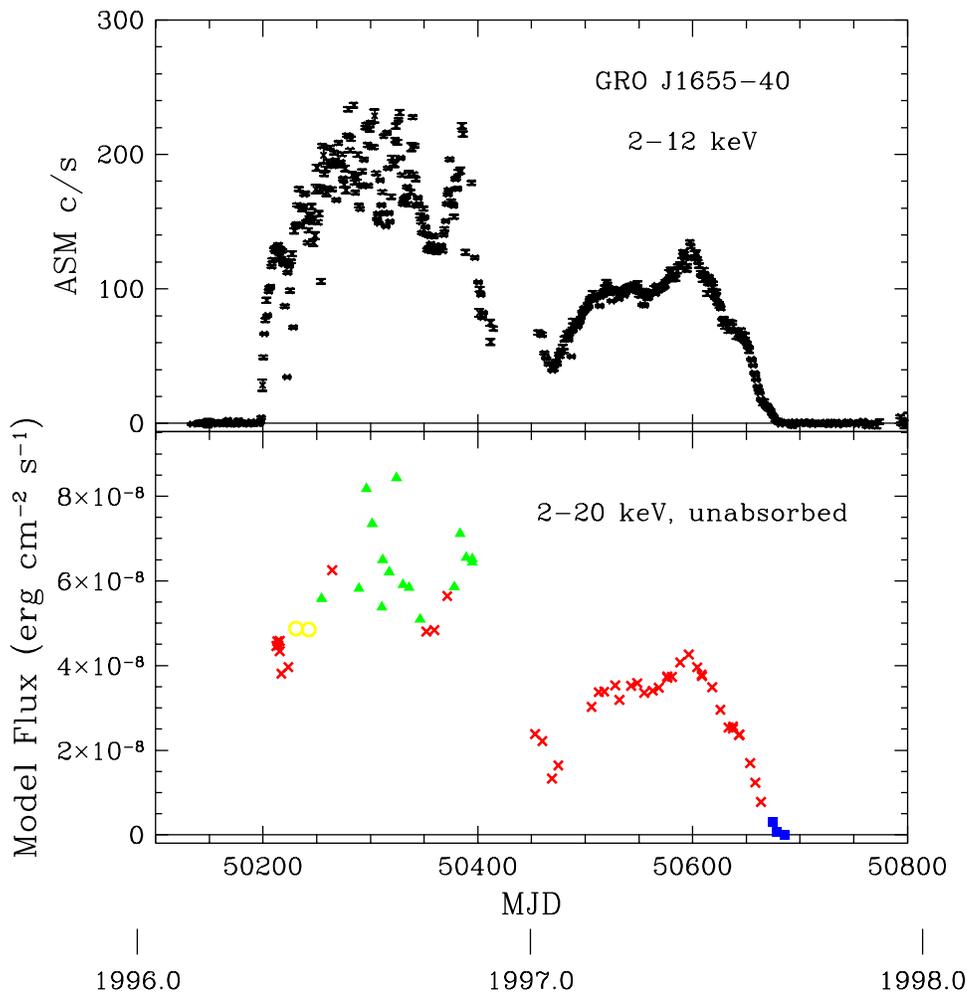}
\caption{X-ray state evolution during the 1996-1997 outburst of
GRO~J1655--40. The top panel shows the ASM light curve.  The bottom
panel shows the model flux from PCA pointed observations.  Here the
symbol type denotes the X-ray state: thermal (red ``x''), hard (blue
square), steep power-law (green triangle), and any type of
intermediate state (yellow circle). } 
\label{fig:lc1655}
\end{figure*}
 
There are other occasions when the spectrum of an outbursting
black-hole binary exhibits substantial non-thermal emission in the form
of an X-ray power-law component. Observations with {\it CGRO}--Ossie were
particularly valuable in showing that there were two types of
non-thermal spectra \cite{gro98}, and this perspective has been confirmed
with {\it RXTE}.  Spectral fits yield
two characteristic values for the photon index. There is a hard state with $\Gamma
\sim 1.7$, usually with an exponential decrease beyond $\sim 100$ keV),
and there is a steep power law ($\Gamma \sim 2.5$), with no apparent cutoff.  
In each case, the corresponding PDS also show distinct differences, relative to
the PDS in the thermal state. 
 
\begin{figure*}[t]
\centering
\includegraphics[width=135mm]{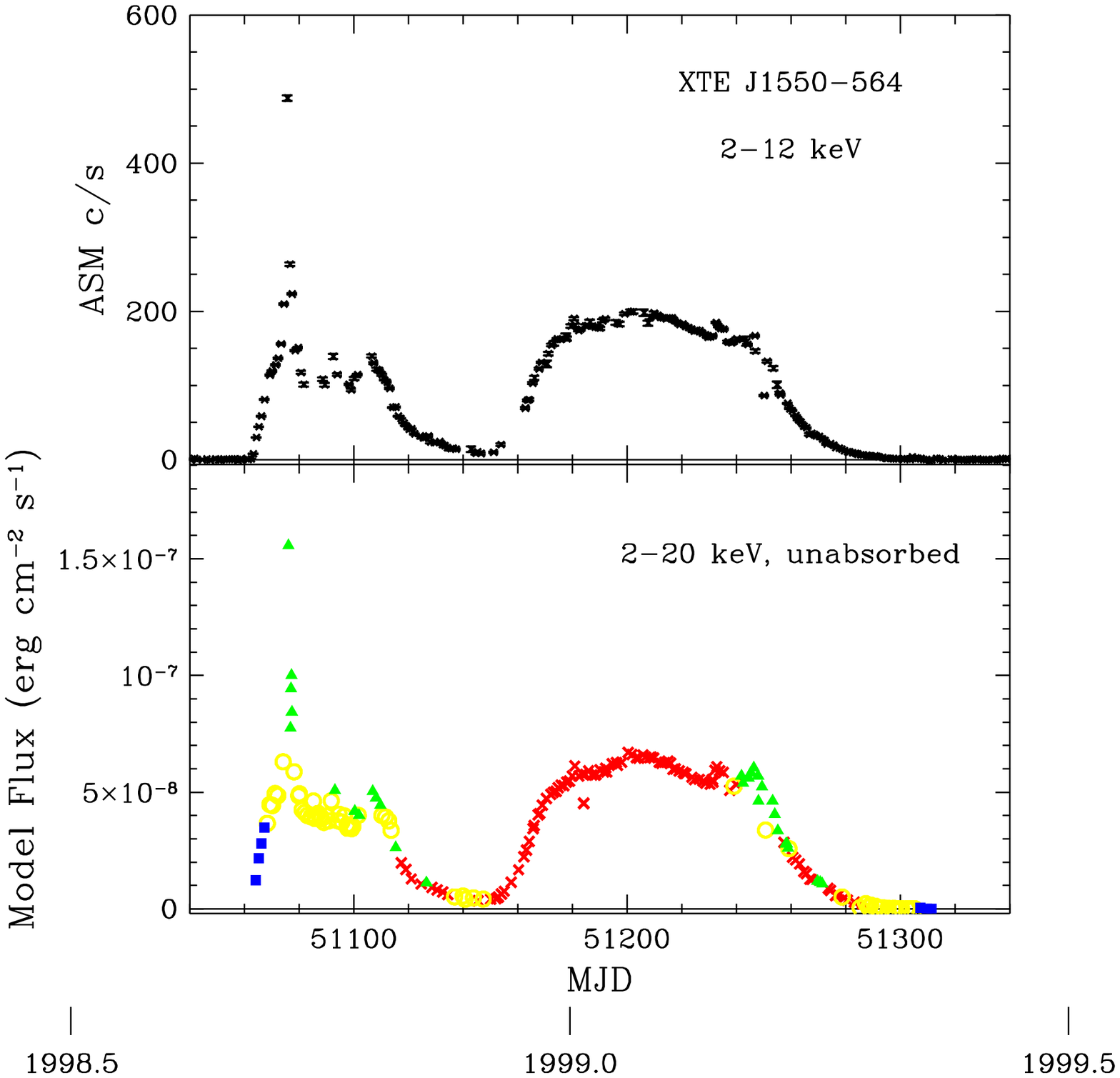}
\caption{X-ray state evolution during the 1998-1999 outburst of
XTE~J1550--564. The top panel shows the ASM light curve.  The bottom
panel shows the model flux from PCA pointed observations, again 
using the symbol type to denote the X-ray state, as defined for 
Fig.~\ref{fig:lc1655}.} 
\label{fig:lc1550}
\end{figure*}

The {\bf hard state} has been clearly associated with the presence of
a steady type of radio jet \cite{gal03,fen05}. Considerations hard states of
for several black-hole binaries allows a definition that is again
based on three X-ray conditions: (1) $f < 0.2$, i.e. the power-law
contributes at least 80\% of the unabsorbed 2--20 keV flux, (2) $1.5 <
\Gamma < 2.1$, for a power-law model, cutoff power law, or broken 
power-law (using $\Gamma_1$), as appropriate, and (3) the PDS yields 
$r > 0.1$. In the hard X-ray state, the accretion-disk component is either
absent or it is modified in the sense of appearing comparatively cool
and large.

The {\bf steep power law} (SPL) state is linked to the strength and
properties of the non-thermal spectrum with $\Gamma \sim 2.5$. QPOs
are frequently seen when the flux from the SPL begins to compete with
thermal component \cite{sob00a}.  The SPL state re-defines the ``very
high'' state, which was brought into use when black hole QPOs were first
discovered and believed to rare and confined to high
luminosity \cite{miy91,miy93}. {\it CGRO} observations have shown that
the SPL may extend to 800 keV or higher \cite{gro98,tom99}. This
forces consideration of non-thermal Comptonization models
\cite{gie03,zdz04}. The QPOs impose additional requirements for an
oscillation mechanism that must be intimately tied to the electron
acceleration mechanism, since the QPOs are fairly coherent ($\nu /
\Delta \nu \sim 12$ ; \cite{rem02a}) and strongest above 6 keV,
i.e. above the limit of the thermal spectrum, which remains visible
during the SPL state. The SPL is further distinguished by its
prevalence when high-frequency QPOs are seen (7 sources; 100-450 Hz),
and the SPL tends to dominate black-hole binary spectra as the
luminosity approaches the Eddington limit \cite{mcc05,rem02}. The SPL
state is defined by: (1) $\Gamma > 2.4$, (2) $r < 0.15$, and (3)
either $f < 0.8$ while a QPO (0.1 to 30 Hz) is present in the PDS
(with $a > 0.01$), \textbf{or} $f < 0.5$ with no QPOs.

\section{Temporal Evolution of X-ray States}

Samples of the energy spectra and PDS for the three states of active
accretion are illustrated for many black hole binaries and candidates
by McClintock and Remillard \cite{mcc05}. Here, the temporal evolution
of states and the corresponding luminosities are shown for three
cases.  Our data selections for {\it RXTE} pointed observations
exclude results when the average flux is below 2 mCrab (or $5 \times
10^{-11}$ erg cm$^{-2}$), since contamination by faint sources in the
Galactic plane within the $1^{\circ}$ PCA field of view may then skew
the spectral parameters derived for the black hole binary.

\begin{figure*}[t]
\centering
\includegraphics[width=135mm]{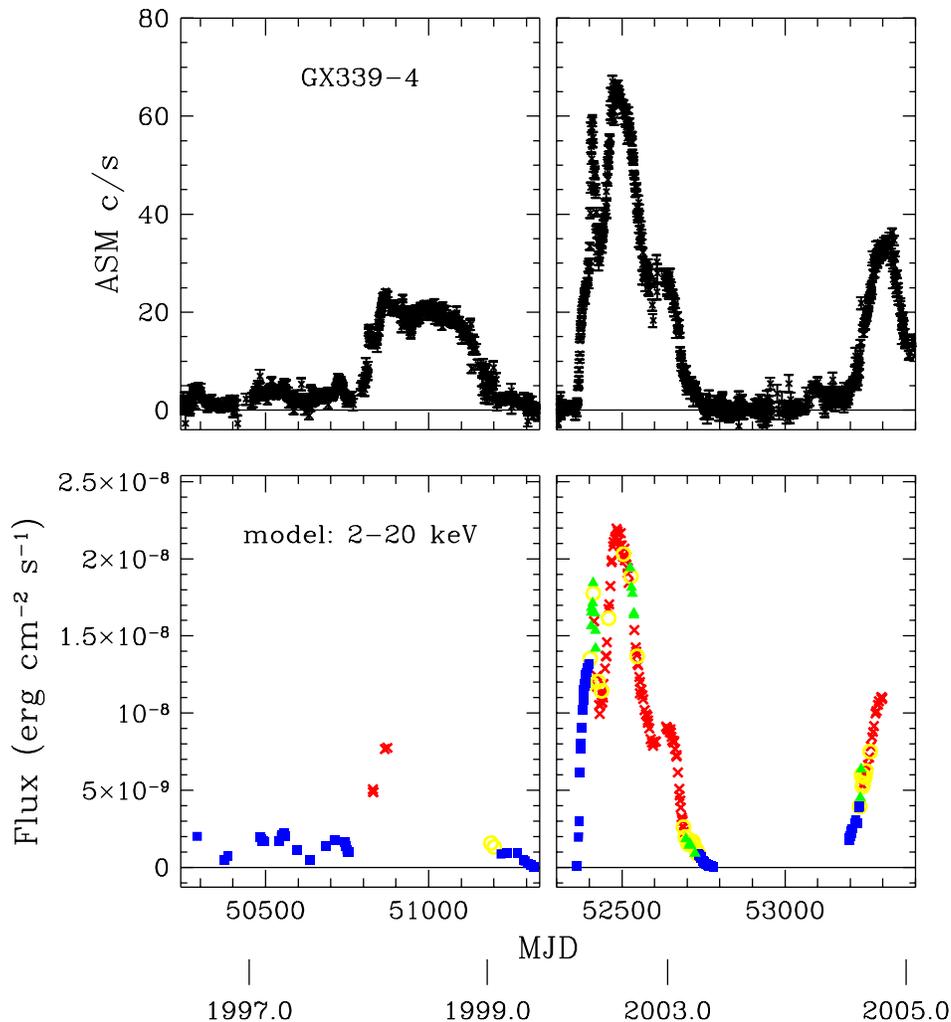}
\caption{X-ray state evolution during multiple outburst of
GX339--4. The top panel shows the ASM light curve.  The bottom
panel shows the model flux from PCA pointed observations, again 
using the symbol type to denote the X-ray state, as defined for 
Fig.~\ref{fig:lc1655}. GX339-4 was in a quiescent state during 
2000 and 2001.}  \label{fig:lcgx339}
\end{figure*}

The temporal evolution of X-ray states for the case of GRO~J1655--40
(1996-1997 outburst) is shown in Fig.\ref{fig:lc1655}. The ASM light
curve is displayed in the top panel, while the unabsorbed flux derived
from spectral fits to {\it RXTE} pointed observations \cite{sob99}
are shown in the bottom panel. The  plotting symbol is used
to represent the X-ray state: thermal (red x), hard (blue
square), SPL (green triangle), and any intermediate type
(yellow circle). This outburst of GRO~J1655-40 is almost entirely
confined to the softer X-ray states (43 thermal cases and 16 SPLs),
while 3 hard state observations are recorded as the source nears
quiescence.  The two intermediate cases happen to be very similar to
each other, and their properties lie just beyond the boundaries for
the thermal state.  The disk fraction in the 2-20 keV flux is high ($f
\sim 0.83$), and there are no QPOs, but the integrated power continuum
is slightly elevated, with $r \sim 0.07$.

\begin{figure*}[t]
\centering
\includegraphics[width=135mm]{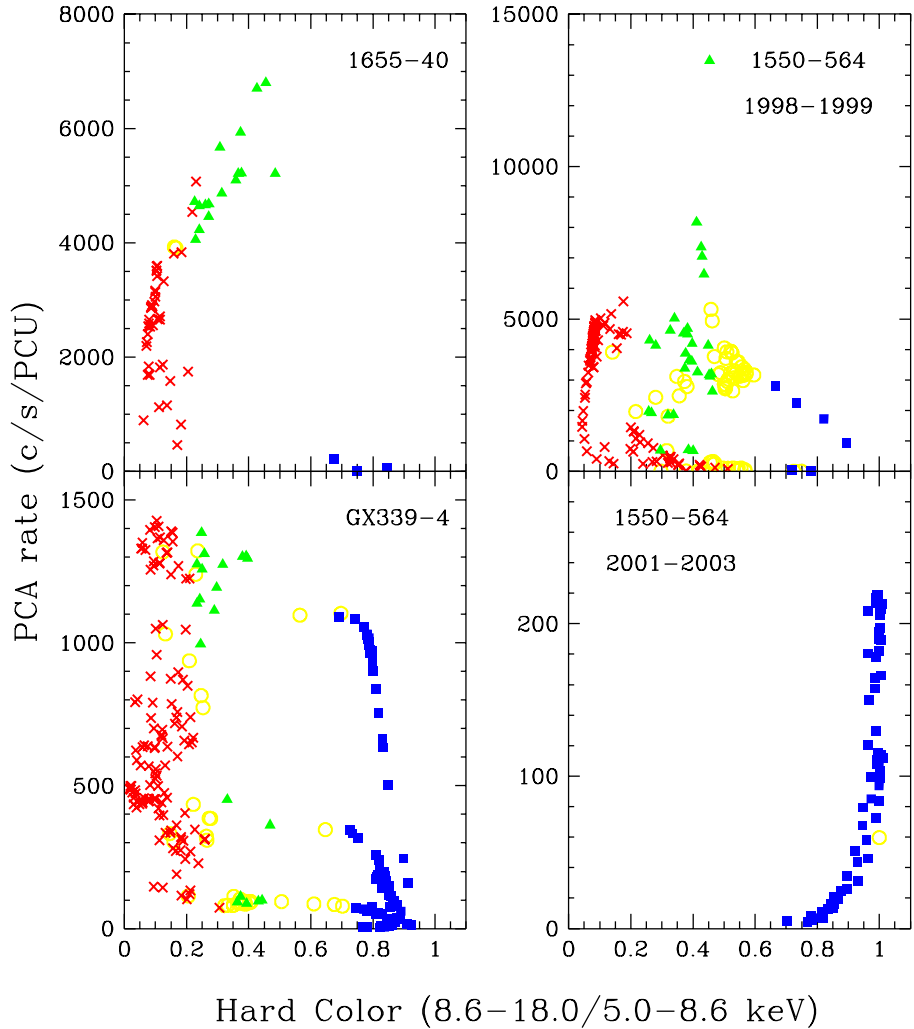}
\caption{X-ray states displayed on the hardness-intensity diagram.
The top-left panel shows GRO~J1655--40 results that correspond with
the light curve shown in Fig.~\ref{fig:lc1655}. Also shown are the
results for the 1998-1999 outburst of XTE~J1550--564 (top-right;
see Fig.~\ref{fig:lc1550}), and all of the observations of GX339--4
(bottom-left; see Fig.~\ref{fig:lcgx339}).  The bottom-right panel
shows the superposition of three fainter outbursts from
XTE~J1550--564 to illustrate the point that some outbursts are
confined to the hard state.}
\label{fig:colint3}
\end{figure*}

The temporal evolution of X-ray states for XTE~J1550--564 (1998-1999
outburst) is shown in Fig.~\ref{fig:lc1550}, using the same format and
X-ray states representations as in Fig.~\ref{fig:lc1655}.  For this
source, spectral modeling efforts follow Sobczak et~al. \cite{sob00b},
except that a broken power law model is used when it improves the fit
significantly, and this happens frequently for observations before MJD
51140.  This outburst exhibits several state transitions, and the
thermal state is the most common condition (106 cases). The SPL state
(30 cases) is again seen when the source reaches highest luminosity;
however, the distribution in luminosity is broad (green triangles,
Fig.~\ref{fig:lc1550}). All three states are seen in the range of 1--4
$\times 10^{-8}$ erg cm$^{-2}$ s$^{-1}$.  There are many observations
(60) of intermediate conditions (yellow circles) during this outburst.
Those in the MJD range 51078--51113 display values for the photon
index and rms power that lie between the SPL and hard states (see
\cite{mcc05} for more detailed discussions), and they coincide with
the appearance of ``C'' type QPOs that vary substantially in frequency
((0.1--10 Hz) \cite{rem02a}.  Near MJD 51300 there is a different
group of intermediate cases that lie between the thermal and hard
states; they show a hard photon index but there are elevated values
of $f$ contributed by the decaying thermal component. This 1998-1999
outburst of XTE~J1550--564 was followed by successively weaker
outbursts (not shown) in 2000, 2001, 2002, and 2003.  The outburst of
2000 again shows multi-state spectral evolution, but the three
subsequent and weaker outbursts appear entirely constrained to the
hard state (see below).

GX339-4 is chosen as a third example to display (see
Fig.~\ref{fig:lcgx339}) the temporal evolution of X-ray states in a
black-hole binary system.  This source is particularly interesting for
two reasons: the relatively high accretion rate from the companion
star produces frequent X-ray outbursts, and the source is known for
extended hard-state episodes, making it a favorite target (along with
Cyg X-1) for radio studies and investigations of the disk:jet
connection \cite{fen99,bel99,cor00,rev01,gal03,cor04}.  Three
outbursts and a long interval in the hard state (1997) are shown in
Fig.~\ref{fig:lcgx339}. GX339--4 was in a quiescent state during 2000
and 2001.  All three active X-ray states are represented by GX339-4:
123 thermal states, 95 hard states, 19 SPLs, and 37 that are
intermediate.

\section{The Unified Model for Radio Jets}

\begin{figure}
\includegraphics[width=55mm]{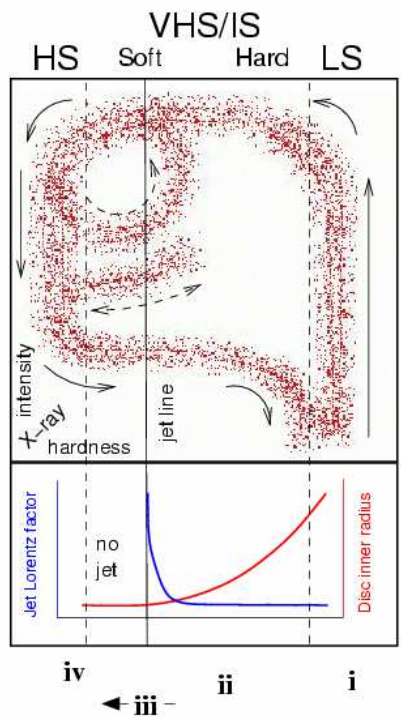}
\caption{A schematic of the model for disk-jet coupling in black hole
binaries from Fender, Belloni, \& Gallo (2004).  The top panel
represents an X-ray HID; the bottom panel illustrates variations in the
jet's bulk Lorentz factor vs. X-ray spectral hardness.  The X-ray
states are labelled at the top using the older terminology:
``high/soft'' (HS; ``thermal state'' in this paper) and ``low/hard''
(LS; ``hard state'' here).  The ``very high/intermediate'' state
(VHS/IS) in this schematic is resolved into the ``SPL'' and different
types of intermediate states in the definitions of McClintock \&
Remillard (2005).}
\label{fig:fender}
\end{figure}

Many researchers choose to investigate the spectral evolution of
accreting black holes in terms of a hardness-intensity diagrams
(HID), where changes in X-ray brightness are tracked vs. a simple ``hardness
ratio'', i.e. the ratio of detector counts in two energy bands
\cite{vdk05,hom01}. The HID is also the format
chosen for the ``unified model for radio jets'' proposed by Fender,
Belloni, \& Gallo \cite{fen04}. 

Having defined the X-ray states and their temporal evolution for
GRO~J1655-40, XTE~J1550-564, and GX339-4 in the preceding figures, we
now display these same results as HIDs.  These data occupy three of
the four panels of Fig.~\ref{fig:colint3}. The horizontal scale of
this plot utilizes the hard color ($HC$) defined by Muno, Remillard,
\& Chakrabarty \cite{mun02} with a normalization scheme that
compensates for PCA gain adjustments and detector evolution that occur
during the {\it RXTE} mission.  PCA spectra of the Crab Nebula yield
values at $HC = 0.68$, with a normalized intensity of 2500 c s$^{-1}$
PCU$^{-1}$. In Fig.~\ref{fig:colint3}, each observation is plotted in
the HID with a symbol type and color that shows the X-ray state, using
the same conventions of earlier figures.

\begin{figure}
\includegraphics[width=80mm]{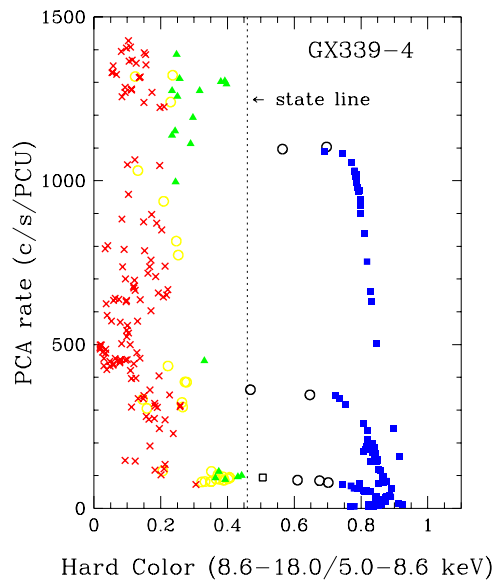}
\caption{The HID of GX339-4, re-displayed after sorting out the
intermediate states into three groups: thermal--SPL intermediates
(29 yellow circles), hard--SPL intermediates (7 black circles), and
hard-thermal (1 black open square).  The dashed line shows the
hardness limit for the SPL state in GX339-4, $HC = 0.46$, and this
line is hypothesized to coincide with the jet line in the schematic
for the unified model of radio jets (Fig.~\ref{fig:fender}).  Similar
$HC$ limits for the SPL states of GRO~J1655-40 and XTE~J1550-564
are evident in Fig.~\ref{fig:colint3}.}
\label{fig:gx339hid}
\end{figure}

The fourth panel (bottom-right) of Fig.~\ref{fig:colint3} shows a
superposition of color:intensity:state results for XTE~J1550-564
during the three weak outbursts noted in \S 2.  For this one panel,
there is a slight change in the definition of the hard state.  The
lower limit for the spectral index is shifted by 0.1, allowing $1.4 <
\Gamma < 2.1$, while keeping other criteria the same.  This
modification is needed to avoid an artificial exclusion of about half
of these observations ($HC > 0.98$) from the hard state.

Fig.~\ref{fig:colint3} again shows that some
outbursts can be dominated by thermal and SPL states
(e.g. GRO~J1655-40, and additional sources, such as 4U~1543-47 and
XTE~J2012+381).  On the other hand, some outbursts (XTE~J1550-564
2001-2003, and also XTE~1118+480 and GS~1324-64) are locked in the hard
state. Complex, multi-state outbursts are seen in
XTE~J1550-564 (1998 and 2000), GX339-4, and sources such as
XTE~J1859+226, 4U~1630-47, and H1743-322.

We wish to compare these state-represented HIDs with the 'unified
model for jets' \cite{fen04}.  The schematic representation
of this model is reproduced here for convenience, and further
discussions by Fender are available in these proceedings.

GX339-4 obviously plays an important role in shaping the dynamics
illustrated in the unified model for jets.  However, as recognized by
the authors\cite{fen04}, and as shown in Fig.~\ref{fig:colint3},
a black hole binary in a given outburst may follow a track that
may be more restricted or more complicated than the tracks shown 
in the model.

The schematic for the disk-jet connection describes not only the flow
of states (in the older state conventions), but it also specifies a
``jet line'' (vertical, solid line center-left in HID of
Fig.~\ref{fig:fender}.  The jet line represents the boundary between
the presence and absence of a steady radio jet, and it also marks a
line of conditions where ballistic ejections are most probable.  Can
such a vertical line be recognized in terms of the revised definitions
of X-ray states?  The answer appears to be 'yes'.  Corbel et
al. \cite{cor04} have investigated the radio properties of some SPL
and intermediate-state observations of XTE~J1650-500 and other sources
and suggest that the jet terminates at the boundary between the
intermediate state and the SPL. In Fig.~\ref{fig:colint3}, this
transition is shown to have a distinct location on the HID, as the
maximum $HC$ value for the SPL state is in the range 0.45--0.48 for
each of the three black-hole binary systems.  We support the
suggestion that this state transition line corresponds to the jet line
and encourage further studies with X-ray and radio data for many
sources.

The hypothesis that the hard end of the SPL state represents the jet
line is illustrated in Fig.~\ref{fig:gx339hid}. Here, the data for
GX339-4 (bottom-left panel of Fig.~\ref{fig:colint3}) is redisplayed
with two changes.  The 37 intermediate designations are resolved into
3 groups, to best describe the nearest two states for a given case.
There are 29 observations with properties that lie between the thermal
and SPL states.  They continue to be plotted as yellow circles in
Fig.~\ref{fig:gx339hid}.  All of these have photon index $\Gamma >
2.4$; fifteen have thermal-like values of $r$ but exhibit either weak
QPOs or relatively low values of $f$; fourteen have SPL-like values of
$r$ and $f$, but there are no QPOs (although some have broad power
features in their PDS). The second group (7 cases) appear to be
intermediate between the hard and SPL states, since they have low $f$
values (i.e. nonthermal spectra) with $2.1 < \Gamma < 2.4$ (i.e. 
photon indices between the hard and SPL states).
This group is plotted with black circles in Fig.~\ref{fig:gx339hid}.
The single, remaining case is considered to be intermediate between
hard and thermal states, since $\Gamma < 2.1$ but $f = 0.34$ ; this is
plotted with a black open square.

\begin{figure*}[t]
\centering
\includegraphics[width=135mm]{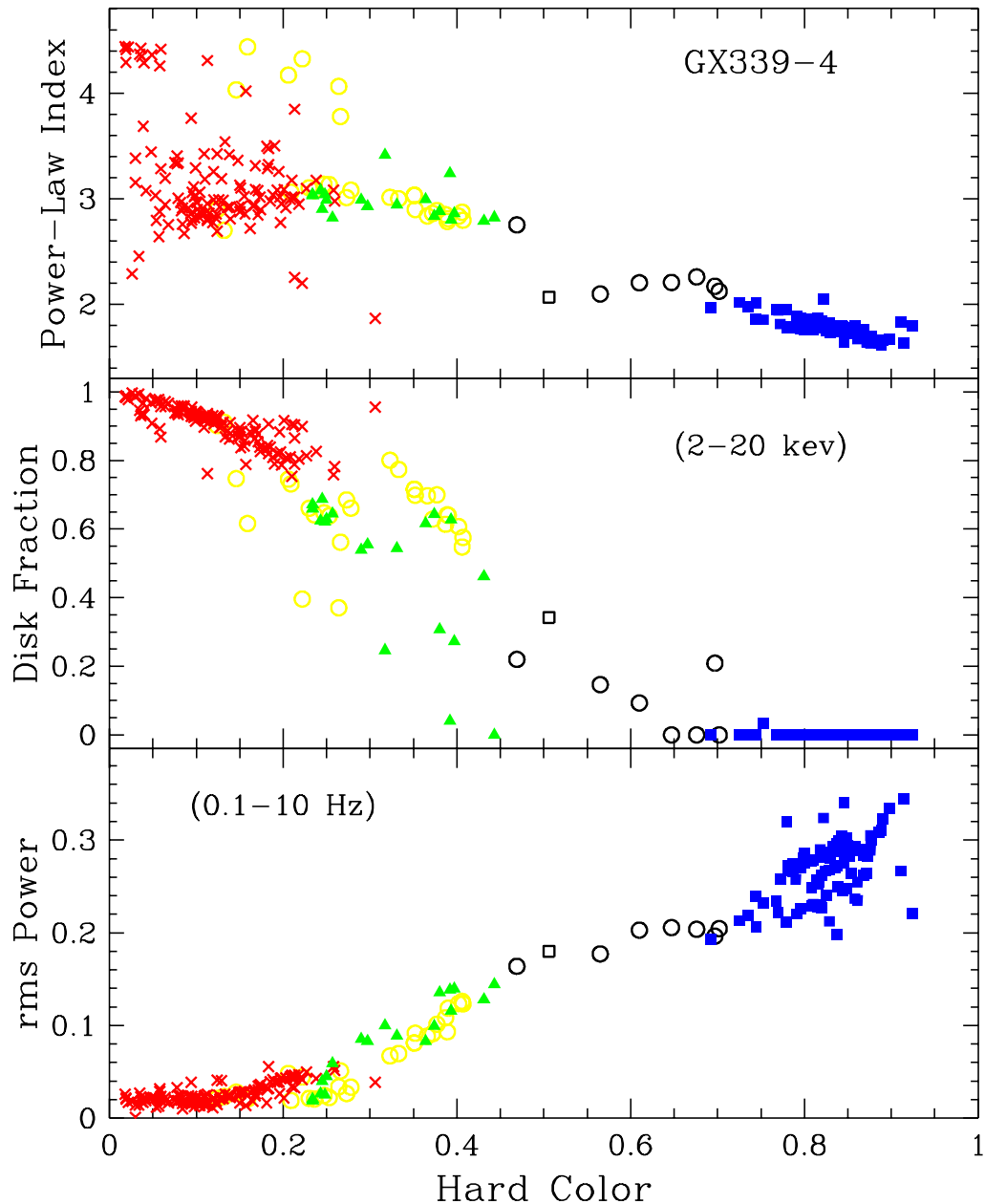}
\caption{Variations of state-defining parameters vs. hard color for
GX339-4. Both the photon index ($\Gamma$) and the integrated rms power
(0.1--10 Hz) change substantially near $HC \sim 0.46$, supporting the
suggestion that this value represents the ``jet line''.}
\label{fig:gx339hr}
\end{figure*}

The effort to synthesize our definitions of X-ray states with the
schematic model for radio jets is encouraging. The identity of the SPL
state and the distinctions between different types of intermediate
states are well preserved on the HIDs. However, no new insights have
been gained, thus far, that are directly pertinent to our
understanding of the jet mechanism. Such interests can be
served by studying the parameters that define X-ray states in greater
detail.

\section{X-ray Parameters vs. Hard Color: Implications for Jets}

There are three continuous X-ray parameters (apart from ``presence of QPOs'') 
that are used to define X-ray states \cite{mcc05}.  These are plotted
vs. hard color in Fig.~\ref{fig:gx339hr} to further investigate
how X-ray properties vary across state transitions and across any
hypothetical jet line.  The symbol choice represents the X-ray state,
as defined for Fig.~\ref{fig:gx339hid}.

As one proceeds from right to left in Fig.~\ref{fig:gx339hr} (i.e. jet
``on'' towards jet ``off''), there do appear to be substantial changes
in both the photon index (top panel) and the rms power continuum
(bottom panel) at the point where the intermediate (hard/SPL)
conditions shift into the SPL state ($HC \sim 0.46$). The disk
fraction in the 2--20 keV band (middle panel) appears to be a more
continuous function of $HC$, and the X-ray appearance of the disk
above 2 keV begins where the hard state (blue squares) 
transitions to hard--SPL intermediate conditions (black circles).

\begin{figure*}[t]
\centering
\includegraphics[width=135mm]{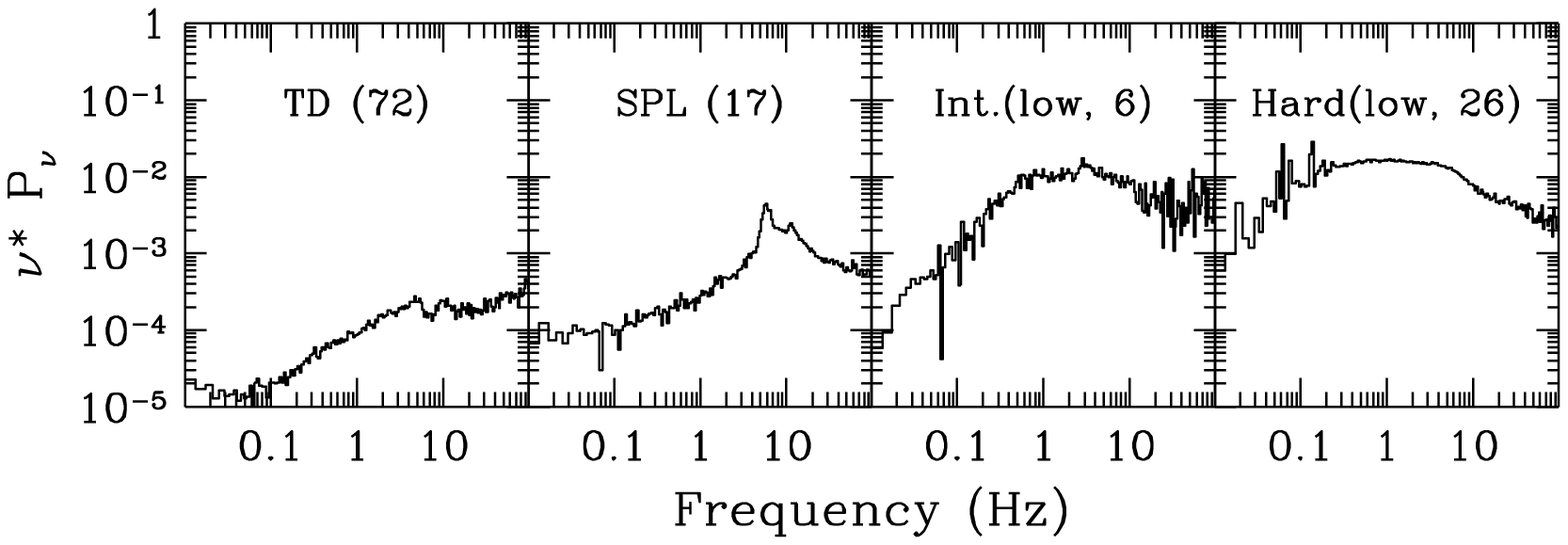}
\caption{Representative PDS for different X-ray states of GX339-4,
displayed in terms of log($\nu \times P_{\nu}$) vs. log($\nu$).  This
yields units of $(rms / \mu)^2$ vs.Hz.  In each state, the averages
are computed for a subset of the observations in a selected range of
$HC$ (see text).  The broad power peak near 1 Hz is here argued to be
an X-ray timing signature associated with the steady radio jet. }
\label{fig:gx339pds}
\end{figure*}

The disappearance of the steady radio jet during ``hard to soft''
X-ray spectral transitions is well known \cite{fen99,cor00,fen05}, and
the behavior of the photon index (top panel of Fig.~\ref{fig:gx339hr})
does suggest that this occurs near $HC \sim 0.46$. However, less
attention has been paid to changes in the power spectrum across this
transition.  The integrated power in the PDS (bottom panel of
Fig.~\ref{fig:gx339hr}) is clearly correlated with the value of $HC$
and the X-ray state.  What is potentially most interesting about this
relationship is the frequency composition of the excess power that
follows the evolution of the source in and out of the hard state.

Representative, average PDS were computed for the different states of
GX339-4 using 72 thermal-state observations with $0.10 < HC < 0.24$
and 17 SPL observations with $0.24 < HC < 0.45$.  For the intermediate
and hard states, we find PDS maxima that are a function of brightness,
and there are low-frequency QPOs in some of the brighter hard states.
These topics needs to be investigated further.  For the present
purpose of illustrating state-specific PDS in GX339-4, we select
intermediate and hard-state observations with lower X-ray intensity,
25--100 c s$^{-1}$ PCU$^{-1}$, to limit the secular variations with
brightness and to represent these states at luminosity levels
typically seen in other sources.  Within these brightness limits, we
select 6 intermediate states with $0.50 < HC < 0.72$ and 26 hard
states with $0.75 < HC < 0.90$. The PDS (weighted averages) for the four states
are displayed in Fig.~\ref{fig:gx339pds}.  

In the SPL state, there is a QPO near 5.8 Hz ($a = 0.03$ in the full
PCA band), with another peak ($a_2 = 0.01$) at the first harmonic.
Note, however, that the average SPL results smear over QPOs with
different amplitudes and frequencies, which are better investigated
using individual SPL observations. 

The PDS for the intermediate (SPL/hard) and hard states appear quite
different; they exhibit broad power peaks \cite{vdk05} with much
higher amplitude. These features can be treated as a broad Lorentzian
peak superposed on a much broader power continuum, i.e. a quadratic
function in log($P_\nu$) vs. log($\nu$) with a fit range of 0.01-1000
Hz.  With this type of PDS model, the intermediate (SPL/hard) state
yields a broad power peak that is centered at $\nu_0 = 0.53$ Hz with
very low coherence parameter: $Q = \nu_0 / FWHM \sim 0.1$.  Most of
the PDS power ($r \sim 0.2$) is taken up in this broad feature.
The hard state PDS can be interpreted with a stronger broad peak 
centered near 0.2 Hz with $Q \sim 0.14$ and $r \sim 0.3$.  

X-ray astronomers have been studying PDS with ``band-limited noise''
or broad power peaks in detail for many years
\cite{wij99,psa99,bel02,pot03,vdk05}.  What is perhaps new, here, is
the demonstrated need to interpret these features as an X-ray
signature that is fundamentally connected to the jet mechanism. These
broad power peaks are strongest in the hard state and they seem to
form only on the right side of the jet line, i.e. the conditions
associated with quasi-steady radio emission.

Pottschmidt et al \cite{pot03} conducted a detailed analysis of
multiple broad peaks with Lorentzian profiles in the PDS of Cyg X-1 in
the hard and intermediate states.  It was shown that the PDS are best
modeled as a superposition of three strong and broad features, e.g
near 0.2, 2, and 6 Hz, with correlated variations in frequency and
with the occasional presence of a fourth peak near 40 Hz.  It was
further shown that the stability of the hard state against transitions
to intermediate-type flares is related to the strength of the third
Lorentzian. These considerations show that the Fourier components
related to the hard state are more complicated than the simplified
results shown here.  However, the point needs to be made that all of
these broad power peaks tend to disappear with the hard power-law
spectrum when an accreting black hole crosses the jet line.

One can imagine the origin of these broad power peaks in very
different ways.  In Keplerian terms, the wide range of excess power
(i.e. 0.01 to 20 Hz) corresponds to gravitational radii of 30--5000 $r_g$
for a 10 \msun black hole with low values of the spin parameter.
If some instability at these radii creates the excess PDS power while
it furnishes the corona that supplies plasma for the jet, then
the base of the jet would be unexpectedly large (0.5 light seconds across
the maximum diameter). This picture would seem to be ruled out by
the high coherence and ms phase lags measured for Cyg X-1 in the hard
and intermediate states (e.g. \cite{pot03}).

Since magnetic fields are widely believed to play a central role in
the formation of jets in black-hole systems, it may be more productive
to consider the PDS power peaks in terms of wave phenomena in a
magnetized disk.  It has been argued that a vertical magnetic field
can excite an ``accretion-ejection'' instability \cite{tag99} leading
to magnetic spiral waves that may produce both QPOs (at the
co-rotation radius for the spiral wave and Keplerian flow in the disk)
and Alfven waves to excite the corona.  The broad power peaks in the
PDS could be seen as another observational challenge for this model
and for MHD simulations of magnetized accretion disks.

Correlations between QPO frequencies and the frequencies associated
with broad power peaks have been demonstrated for hard and
intermediate state observations of both black-hole binaries and
accreting neutron stars \cite{wij99,psa99,bel02}. The neutron-star
connection naturally motivates the presumption that the behavior of
X-ray power peaks is dictated by standard accretion-disk physics.
However, the consideration of black-hole hard states and the jet line
motivates an alternative viewpoint: the frequency relationships for
broad power peaks are primarily exhibiting the behavior of non-thermal
processes that are related to the jet mechanism.  As a
corollary question, we may need to understand why jets appears to be
less efficient in the hard states of accreting neutron stars
\cite{mig03,mun05}.

\begin{acknowledgments}
  This work was supported by the NASA contract to MIT for the ASM and
  EDS instruments on {\it RXTE}. Special thanks are extended to Jeff
  McClintock for many contributions to this research.
\end{acknowledgments}

\bigskip % extra skip inserted

\end{document}